# Minimizing Uplink Delay in Delay-Sensitive 5G CRAN platforms

Ali Ataie
*Electrical Engineering department*
*Sharif university of Technology*
Tehran, Iran
ali.ataie@ee.sharif.edu

Borna Kanaanian
*Electrical Engineering department*
*Sharif university of Technology*
Tehran, Iran
borna@ee.sharif.edu

Babak H. Khalaj
*Electrical Engineering department*
*Sharif university of Technology*
Tehran, Iran
khalaj@ee.sharif.edu

*Abstract*— In this paper, we consider the problem of minimizing the uplink delays of users in a 5G cellular network. Such cellular network is based on a Cloud Radio Access Network (CRAN) architecture with limited fronthaul capacity, where our goal is to minimize delays of all users through an optimal resource allocation. Earlier works minimize *average* delay of each user assuming same transmit power for all users. Combining Pareto optimization and Markov Decision Process (MDP), we show that every desired balance in the trade-off among *infinite-horizon average-reward* delays, is achievable by minimizing a properly weighted sum delays. In addition, we solve the problem in two realistic scenarios; considering both power control and different (random) service times for the users. In the latter scenario, we are able to define and minimize the more preferred criterion of total delay vs. average delay for each user. We will show that the resulting problem is equivalent to a *discounted-reward infinite-horizon* MDP. Simulations show significant improvement in terms of wider stability region for arrival rates in power-controlled scenario and considerably reduced sum of users' total delays in the case of random service times.

## I. Introduction

Improving spectral efficiency (SE) is of key importance in 5G networks in order to satisfy demanding QoS requirements such as minimal experienced delay for users. It is well-known that network-wide BS cooperation methods, e.g., Joint Transmission Coordinated Multi Point (JT CoMP) can increase SE. While JT CoMP can be implemented in either a centralized or a distributed manner [4], the former is of higher interest due to its cost efficiency. By centralizing the processes in a base-band process unit (BPU), the new nodes will be in the form of lower complexity Radio-Remote Heads (RRHs). Such structure is the core idea of a Cloud Radio Access Network (CRAN) in 5G. But the cost of cheaper BSs leads to high-throughput "Fronthaul network" between RRHs and the BPU. Moreover, in a large network it may not be economical or even practical to directly connect all the RRHs to the BPU. Instead, a limited-capacity network mediates between the BPU and the distant RRHs. Optimal fronthaul allocation for maximizing network throughput in uplink is investigated in [3]. It is shown in [7] that assuming a Gaussian distribution for the input signal as well as a Gaussian quantization noise independent from the input, leads to optimality of Gaussian quantization for the problem in [3].

In this paper, our goal is to minimize the users' delays in uplink of a CRAN by efficient allocation of the resources; the fronthaul capacity and the users' transmit powers. In [5] and [6], in order to minimize the time average of the delay for each user, a long-term optimization, rather than a one-shot approach is used. Authors in [5] and [6] utilize Markov Decision Process (MDP) framework to formulate such long-term optimization and model the problem as an infinite-horizon average-reward MDP.

Moreover, minimizing average delays of all the users is a multi-objective optimization, with trade-offs among users. Weighted sum of users' average delays is considered in [6] as the objective function in order to achieve various balances for the mentioned trade-offs. Combining *Pareto optimization* and *MDP theories*, we prove that all the possible balances for the trade-offs among users are achievable by optimizing a properly "weighted sum of average delays".

Papers [5, 6, and 12] considered the same transmit power for all users. In an illustrating example, we show that a fronthaul allocation integrated with uplink power control, can reduce the average delays of users. We use an approximation similar to [6] to find the jointly optimal policy analytically, using the perturbation theory. Moreover, previous works such as [5, 6, and 12], target to minimize the *time-average* of delay for each user, which necessitates considering a sufficiently long common service duration for all the users. But in a real network, users ask for services of different durations. Therefore, we will also consider this scenario and try to minimize users' delays in the case of finite random service durations. With a finite horizon, one is able to target minimization of *total delay*, instead of average delay, for each user as a higher QoS criterion.

This paper is organized as follows: In section II, the system model is described. In Section III, we extend the multi-objective (Pareto) optimization to "infinite-horizon, average reward" objective functions. In section IV, we utilize our proposed model and show that every desired balance in the trade-off among average delays of users is achievable by minimizing a properly weighted sum of the average delays. In section V, we find the jointly optimal policy for allocation of *fronthaul capacity* and *users' transmit powers*. In section VI, we formulate the discounted-reward MDP model for finite-duration services. Simulations presenting the effectiveness of the proposed solutions are presented in section VII. Finally, section VIII concludes the paper.

## II. System Model

We consider a CRAN architecture with *m* RRHs linked to the BPU through an intermediate network such as the one shown in Fig. 1. Here, we assume universal frequency reuse and *m* users, where the number of users can be increased in practice by a desired multiplexing scheme such as FDM.

### A. Wireless channels

Since signals of users are mixed at RRHs due to the universal frequency reuse, decoding of the received signals are done jointly at BPU (JT CoMP). In order to avoid the high complexity of running JT CoMP for all RRHs in a large network, it can be implemented for groups of neighboring RRHs. The group of RRHs for which JT CoMP is utilized, is called a cooperating cluster. The time is slotted into intervals of $\tau$ seconds. In a cooperating cluster of $n$ cells, the received baseband signal by RRH $i$ at time slot $t$, is given by (1).

$$y_i(t) = \sum_j h_{ij}(t) x_j(t) + n_i(t) , \quad (1)$$

where, $h_{ij} \in \mathbb{C}$ is the channel coefficient from user $j$ to RRH $i$. $x_j(t) \in \mathbb{C}$ is the transmitted signal from user $j$ at time slot $t$, and $n_i(t)$ is the complex AWGN noise at the receiver of RRH $i$ with variance $N_0$. We assume $h_{ij}$ as a composition of a path-loss gain $L_{ij} \in \mathbb{R}$ and a small-scale fading term $\tilde{h}_{ij} \in \mathbb{C}$ such that $h_{ij} = \tilde{h}_{ij}\sqrt{L_{ij}}$. While the large-scale fading gain $L_{ij}$ is assumed to be constant due to its slow change rate, $\tilde{h}_{ij}$ is updated at each time slot, according to a Gaussian distribution with zero mean and unit variance. Subsequently, the RRHs transmit a quantized version of the received signal to the BPU.

The quantization precision is bounded by the fronthaul capacity allocated to each RRH. We assume that the quantization noise $n_{q,i}$ is also Gaussian with zero mean and variance given by:

$$N_i = \frac{\sum_{j=1}^{n} \|h_{ij}\|^2 p_j + \sigma^2}{2^{C_i} - 1} . \quad (2)$$

where, $p_j$ is the transmit power of user $j$ and $C_i$ is the fronthaul capacity allocated to RRH $i$ in bits/sec/Hz. Consequently, the quantized version of the received signal by RRH $i$ at BPU is given by:

$$\hat{y}_i(t) = \sum_j H_{ij}(t) x_j(t) + n_i(t) + n_{q,i}(t) . \quad (3)$$

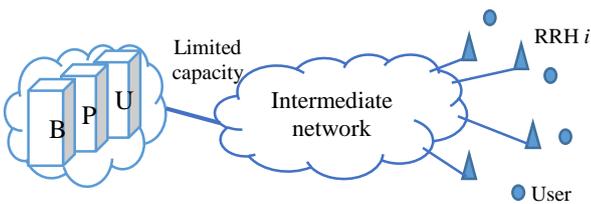

Fig. 1. The considered network architecture

The vector form of (3) is expressed as:

$$\hat{\mathbf{y}} = \mathbf{Hx} + \mathbf{n} + \mathbf{n}_q , \quad (4)$$

Where, $\mathbf{H}$ is the matrix of channel gains, while $\mathbf{n}$ and $\mathbf{n}_q$ are the vectors of the thermal and quantization noises, respectively. The BPU zero forces the received signals $\hat{\mathbf{y}}$ to retrieve each individual user's signal:

$$\hat{\mathbf{x}} = \mathbf{S}\hat{\mathbf{y}} \quad ; \mathbf{S} = \mathbf{H}^{-1} , \quad (5)$$

where, $\hat{\mathbf{x}}$ represents the estimated vector of signals of users within the cooperating cluster. The end-to-end transmission rate of user $i$, $R_i$, is given by:

$$R_i(\mathbf{H},\mathbf{C}) = \frac{W}{2}\log_2\left(1 + \frac{p_i}{\sum_{j=1, j\neq i}^{n} S_{ij}^2 (N_j + N_0)}\right) , \quad (6)$$

which is a function of channel state $\mathbf{H}(t)$ and the fronthaul capacity allocation $\mathbf{C}(t)$, at time slot $t$. In the above equation, $W$ is the bandwidth used by all the users in the cooperating cluster. Time indexes are removed for brevity.

### B. User traffic queues

User $i$ generates $A_i(t)$ bits according to a Poisson distribution at the end of time slot $t$ and stores them in his queue, which contains $Q_i(t)$ bits. At each time slot $t$ and prior to its end, $R_i(t).\tau$ bits are transmitted if there is enough number of bits in the queue and $Q_i(t)$ bits, otherwise. Therefore, the evolution of $Q_i(t)$ is described by

$$Q_i(t+1) = [Q_i(t) - R_i(t).\tau]^+ + A_i(t) , \quad (7)$$

where the operator $[.]^+$ denotes $\max\{.,0\}$.

### C. MDP formulation:

Assuming the same service time for all users, the goal is to minimize the expected time average of the delay experienced by each user. The expectation depends not only on the distribution of wireless channels, but also on how fronthaul capacity is assigned to RRHs and transmit powers to the users. Therefore, we define an infinite-horizon average-reward MDP, with the pair of *wireless channel gains* and *queues' lengths*, as its system state: $\chi(t) = (\mathbf{Q}(t), \mathbf{H}(t))$, which is governed with a policy $\pi$ of decisions $d_t$ at each time slot $t$:

$$d_t = [p_{1,d}(t),...,p_{n,d}(t); C_1(t),...,C_n(t)] , \quad (8)$$

here, $C_i(t)$ is the fronthaul capacity allocated to RRH $i$, and $p_{j,d}(t)$ is the dynamic part of the transmit power assigned to user $j$, which will be described explicitly in section V. Consequently and based on Little's law, the expected average delay for each user $i$ is given by:

$$\bar{D}_i(\pi) = \limsup_{T \to \infty} \frac{1}{T} \sum_{t=1}^{T-1} E^\pi \left[ \frac{Q_i(t)}{\lambda_i} \right]. \quad (9)$$

Moreover, note that such MDP is constrained due to limitations on total fronthaul capacity and the upper limit on the average dynamic power for each user:

$$\sum_i \bar{C}_i \leq C_{tot} \; ; \; \bar{C}_i(\pi) = \limsup_{T \to \infty} \frac{1}{T} \sum_{t=1}^{T} E^\pi [C_i(t)], \quad (10)$$

$$\bar{p}_{i,d}(\pi) = \limsup_{T \to \infty} \frac{1}{T} \sum_{t=1}^{T-1} E^\pi [p_{i,d}(t)] \leq p_{i,d\_max} \quad 1 \leq i \leq n \quad (11)$$

### III. MULTI-OBJECTIVE OPTIMIZATION

Consider the problem of minimizing $n$ correlated infinite-horizon average-reward objective functions with a finite and countable state-space in common:

Problem 1:

$$\min_\pi \bar{D}_1(\pi), \bar{D}_2(\pi), ..., \bar{D}_n(\pi) \quad s.t. \begin{cases} \bar{C}_k(\pi) \leq \alpha_k & k=1,...,r \\ \bar{S}_k(\pi) = \beta_k & k=1,...,l \end{cases};$$

$$\bar{D}_k(\pi) = \limsup_{T \to \infty} \frac{1}{T} \sum_{t=1}^{T} D_k^\pi(t)$$

$$\bar{C}_k(\pi) = \limsup_{T \to \infty} \frac{1}{T} \sum_{t=1}^{T} C_k^\pi(t) \; , \; \bar{S}_k(\pi) = \limsup_{T \to \infty} \frac{1}{T} \sum_{t=1}^{T} S_k^\pi(t)$$

where $\bar{D}_i$s are the objective functions (average rewards or costs), $\bar{C}_k$s are the inequality constraints, and $\bar{S}_k$s are the equality constraints.

**Theorem 1.** ([1], Theorem 3.5, paraphrased): *In a multi-objective optimization, if the objective range is convex, then its Pareto front is identical with the optimal range of the problem of minimizing the weighted sum of functions with weights spanning over all possible non-negative values.*

Based on theorem 1, we have to investigate the convexity of the objective range in Problem 1.

**Lemma 1.** *The objective range of Problem 1, is convex.*
*Proof:*
Consider two points $A = (\bar{D}_1, \bar{D}_2, ..., \bar{D}_n)$ and $B = (\bar{D}'_1, \bar{D}'_2, ..., \bar{D}'_n)$ in the objective range, resulting from policies $\pi_1$ and $\pi_2$, which are not necessarily stationary, respectively. In any infinite-horizon average reward problem, one can find a sufficiently long duration $T_1$ over which:

$$\lim_{T \to \infty} \frac{1}{T} E^{\pi_1} \sum_{t=1}^{T} D_k(t) \approx \frac{1}{T_1} E^{\pi_1} \sum_{t=1}^{T_1} D_k(t) \quad \forall k \in \{1,...,n\}, \quad (13)$$

We can similarly find a duration $T_2$ for point B. Now, consider a sufficiently large time interval, which can be partitioned into a desired numbers of $T_1$ and $T_2$ intervals and use $\pi_1$ and $\pi_2$ on them, respectively. Therefore, all points on the segment 'AB' in the objective space are achievable through time-sharing between $\pi_1$ and $\pi_2$, and hence the objective range is convex. A similar "time sharing" idea is also utilized in [9].

**Theorem 2** ([2], Theorem 8.9.6, part b):
*Suppose that the state-space of the infinite-horizon average-reward constrained MDP is finite and countable and there exists an optimal policy $\pi^* \in \Pi^{HR}$, which is an optimal solution to the constrained MDP. Then, there exists a stationary policy $(d^*)^\infty \in \Pi^{SR}$ which is also optimal.*

Note that $\Pi^{HR}$ and $\Pi^{SR}$ are "history-dependent randomized" and "stationary randomized" policy spaces, respectively.

**Theorem 3:** *An optimal stationary policy for every desired point on the Pareto front of Problem 1 can be found, by minimizing a properly weighted sum of the objective functions.*
*Proof*: Based on Theorem 1 and Lemma 1, every point on the Pareto front of the Problem 1 is also achievable by minimizing a weighted sum of objective functions, with proper weights. Hence, there is a policy for the weighted sum MDP which results in the corresponding point on the Pareto front. Using theorem 2, we deduce the existence of an optimal <u>stationary</u> policy.

### IV. TRADE-OFF AMONG USERS' DELAYS AND PROOF OF SUITABLITY FOR WEIGHTED SUM CRITERION

In minimizing users' average delays as a multi-objective optimization, the key point is to choose an optimization method capable of realizing all possible trade-offs among the users. Pareto front is the mathematical description of the possible trade-offs. Theorem 3 considers MDPs with finite and countable state-spaces, while the state-space of delay minimization problem is not. Explicitly, the *wireless channel gains* take real values in an unlimited range. The set of queue lengths is also unlimited in theory. In practice, however, there are upper limits for both and in fact channel gains can be discretized by quantizing them according to the needed precision for calculations. Additionally in digital implementations of the corresponding algorithm in a CRAN, the mentioned quantization is necessary.

**Corollary 1**: *If the state-space of the "delay minimization problem" is made finite and countable by bounding and discretizing, then an optimal stationary policy for every desired balance in the trade-off among users can be found, by minimizing a properly weighted sum of users' average delays.*

### V. JOINT OPTIMAL POLICY FOR USERS' POWER CONTROL AND FRONTHAUL ALLOCATION

Following, is an example on how joint allocation can improve delay reduction. Consider a simple cooperating cluster with two RRHs and two users. In this case, by substituting $N_j$ based on (2), the end-to-end rate of user $i$ according to (6) can be written as:

$$R_i \approx .5W \log_2(1 + p_i/(a_i N_0 + 2^{-C_1}(b_i p_1 + d_i p_2) + 2^{-C_2}(f_i p_1 + g_i p_2))) \quad , (14)$$

where $b_i$, $d_i$, $f_i$ and $g_i$ are given by the corresponding terms in (6), and are functions of channel coefficients. If the multipliers of $2^{-C_1}$ and $2^{-C_2}$ be of the same order of magnitude for some channel realizations, then the data rates of the two users stay almost the same for all fronthaul allocations. In such cases, only changing the transmit powers can adjust the data rates in order to reduce delays. Therefore, in addition to the general gains of *power control*, using it in combination with fronthaul allocation can enhance delay performance as required in 5G.

*A.    Formulating the MDP problem*

The transmit power of each user consists of a constant part and a dynamic part:

$$p_i(t) = p_{i,0} + p_{i,d}(t) \quad 1 \leq i \leq n , \quad (15)$$

With the objective functions and the constraints defined in previous parts, we can state the MDP problem as:

$$\min_\pi \sum_{i=1}^n \beta_i \overline{D}_i(\pi) \text{ s.t. } \begin{cases} \sum_{i=1}^n \overline{C}_i(\pi) \leq C_{tot} \\ 0 \leq \overline{p}_{i,d}(\pi) \leq \overline{p}_i - p_{i,0} \quad \forall i: 1 \leq i \leq n \end{cases}, (16)$$

The Lagrangian of this problem is:

$$\mathcal{L} = \limsup_{T \to \infty} \frac{1}{T} \sum_{t=1}^{T-1} E^\pi \left[ \sum_i \left( \beta_i \frac{Q_i(t)}{\lambda_i} + \gamma C_i(t) + \mu_i p_{i,d}(t) \right) \right], (17)$$

Considering $\gamma$ and $\mu_i$s as constants and optimizing $\mathcal{L}$ with respect to $\pi$, we can interpret $\gamma$ as the cost of one $\frac{bit}{sec.Hz}$ of total fronthaul capacity and each $\mu_i$ as the cost of one *watts* of dynamic power for user $i$. Consequently, we can define the cost function at each time-slot as:

$$c(\mathbf{Q}(t), \pi(\chi(t))) = \sum_i \left( \beta_i \frac{Q_i(t)}{\lambda_i} + \gamma C_i(t) + \mu_i p_{i,d}(t) \right)$$
,   (18)

Substituting (18) as the new cost function, theorems 1 and 2 of [6] hold here as well. Common methods of value or policy iteration, do not provide a closed-form answer and impose high computational complexity. Assuming small-enough cross-links $L_{ij}$ in comparison with straight links $L_{ii}$, we utilize the analytical method of perturbation theory as in [6], to find a well-approximated analytical solution for the MDP in (16).

**Theorem 4**: (extension of theorem 2 of [6])

*Define* $\delta = max\{L_{ij} : \forall i \neq j\}$ *and assume there exists a constant* $c^\infty$ *and a function J* ($\mathbf{Q}$; $\delta$) *of class* $\mathbb{C}^2(\mathbb{R}_+^n)$ *which satisfy:*

1) For all $k$, $\frac{\partial J(\mathbf{Q};\delta)}{\partial Q_k}$ *is an increasing function of all* $Q_k$
2) $J(\mathbf{Q}; \delta) = O(\mathbf{Q}^2)$
3) *The following partial differential equation:*

$$E\left[\min_{\pi(\chi)} \left[ \sum_{i=1}^n \left( \beta_i \frac{Q_i}{\lambda_i} + \gamma C_i + \mu_i P_{i,d} \right) - c^\infty + \cdots \right.\right.$$
$$\left.\left. \sum_{i=1}^n \left( \frac{\partial J}{\partial Q_i} (\lambda_i - R_i(\mathbf{H},\mathbf{C})) \right) \right] \middle| \mathbf{Q} \right] = 0 . \quad (19)$$

, *with the boundary condition* $J(0; \delta) = 0$.

Then, the optimal average cost and the *priority function of the data flows are respectively given by* $\theta^* = c^\infty + o(1)$ and $V^*(\mathbf{Q}) = J(\mathbf{Q};\delta) + o(1)$ in which, the error term $o(1)$ approaches asymptotically to zero, as $\tau$ decreases.

*B.    Solution of the MDP problem:*

Similar to [6], we first compute $J(\mathbf{Q}; 0)$ using its decomposable structure and later compute the effect of non-zero cross-links up to their second Taylor term. The approximate priority function $J(\mathbf{Q}; 0)$ can be decomposed to *n* sub-priority functions as:

$$J(\mathbf{Q};0) = \sum_{i=1}^n J_i(Q_i) , \quad (20)$$

The proof mimics that of [8], since the only difference is addition of power terms to the cost function, which acts the same as previously-existing capacity terms.

Based on (19), only the derivatives of the approximate priority function $J$ is required to calculate the optimal policy. The following theorem provides the equation for computing these derivatives.

**Theorem 5**: *The derivative of the i$^{th}$ sub-priority function, $J_i'(Q_i)$ satisfies the following equation:*

$$\beta_i \frac{Q_i}{\lambda_i} + \gamma E[C_i^*] + E[\mu_i P_{i,d}^*] + \frac{\partial J_i(Q_i)}{\partial Q_i}(\lambda_i - E[R_{i0}^*])$$
$$-c_i^\infty = 0 , \quad (21)$$

The closed-form approximations for the expectations in (21), as well as its proof, are presented in Appendix A.

In order to account for the effect of cross-links, we use the Taylor expansion of $J(\mathbf{Q};\delta)$ up to its second term.

$$J(\mathbf{Q};\delta) = J(\mathbf{Q};0) + \sum_{i \neq j} \sum_{j=1}^m L_{ij} \tilde{J}_{ij}(\mathbf{Q}) + O(\delta^2), \quad (22)$$

The functions $\tilde{J}_{ij}(\mathbf{Q})$ may be called the joint priority functions.

**Theorem 6**:
*The derivatives of the joint priority functions with respect to queues' lengths are given by, where, $v_i$s are constants:*

$$\frac{\partial \tilde{J}_{ij}}{\partial Q_i} = \frac{\gamma}{L_{ii}} \frac{v_5(1-v_1)\alpha_i}{(E[R_{i0}^*] - \lambda_i)} \quad , \quad \frac{\partial \tilde{J}_{ij}}{\partial Q_j} = \frac{\gamma}{L_{jj}} \frac{v_1 v_4 \alpha_j}{(E[R_{j0}^*] - \lambda_j)} \quad (23)$$

*Proof:* See appendix B

### D. The optimal allocations:

After computing $J(\mathbf{Q}; \delta)$ (for a certain distribution of wireless channels and set of arrival rates), we can substitute it in the minimization stage of (19) rewritten in (24), to derive the optimal allocations for fronthaul capacity and transmit powers at each system state $\chi(t) = (\mathbf{Q}(t), \mathbf{H}(t))$. Note that the expected effect of future events on this allocation are embedded in the function $J(\mathbf{Q}; \delta)$.

$$\min_{[C_1,\ldots,C_n,P_{1,d},\ldots,P_{n,d}]} \left[ \sum_{i=1}^{n} \left( \beta_i \frac{Q_i}{\lambda_i} + \gamma_i C_i + \mu_i P_{i,d} \right) - c^{\infty} + \ldots \right. \\ \left. \sum_{i=1}^{n} \left( \frac{\partial J(\mathbf{Q}; \delta)}{\partial Q_i} (\lambda_i - R_i(\mathbf{H}, \mathbf{C})) \right) \right]. \quad (24)$$

## VI. DELAY MINIMIZATION FOR FINITE DURATION SERVICES

In a real network, users ask for services of limited duration. Such durations are usually modeled as random variables with exponential distribution (geometric in the case of slotted time). Since total reward is definable in finite-horizon case, in this section we minimize the total cost (delay) for each user. Optimizing the total and average rewards are equivalent for a fixed horizon length, but are quite different in the case of random horizon length. It is shown that optimizing total reward for geometrically distributed horizon lengths, is equivalent to optimizing an infinite-horizon discounted-reward problem. It should be noted that the optimization horizon is limited by the user whose service terminates first. Then, a new optimization begins with the rest of the users. The users are assumed to be independent with a geometrical parameter $\mu_i$ for user $i$. Consequently, the optimization length will follow a geometrical distribution with parameter $\mu = \prod_{i=1}^{L} \mu_i$.

### A. Problem formulation:

First, we shall analyze the effect of random horizon length. The following lemma, addresses this issue:

**Lemma 2** ([2], proposition 5.3.1):
*A total-reward finite-horizon MDP with geometrically distributed random horizon length with parameter $\mu$, is equivalent to a discounted-reward infinite-horizon MDP, with a discount rate of $\mu$.*
*Proof*: see Appendix C

Based on lemma 2, the expected total delay for user $i$, in the case of random finite service durations, can be written as:

$$\overline{D}_i(\pi) = \limsup_{T \to \infty} \sum_{t=1}^{T-1} E^{\pi} \left[ \frac{Q_i(t)}{\lambda_i} \right] \mu^{t-1}, \quad (25)$$

Since this infinite-horizon problem is not average-reward, we cannot utilize Theorem 3. Therefore, here we use the "weighted sum of total delays" as the objective function. The decision rule at time $t$ is to assign fronthaul capacity for each RRH:

$$d_t = [C_1(t), C_2(t), \ldots, C_n(t)], \quad (26)$$

Moreover, we have to update the fronthaul constraint to match the discounted-reward formulation as:

$$\sum_{i=1}^{n} \overline{C}_i(\pi) \leq C_{tot} \; ; \; \overline{C}_i(\pi) = \limsup_{T \to \infty} \sum_{t=1}^{T-1} E^{\pi} [C_i(t)] \mu^{t-1}, \quad (27)$$

This cost can be interpreted as regulation for the total fronthaul capacity used by each RRH which in turn, is more realistic than regulating for the time-average of the consumed fronthaul capacity. The Lagrangian for minimizing (25) constrained to (27) equals:

$$\min_{\pi} \quad L(\pi) = \sum_{i=1}^{L} \left\{ \beta_i \overline{D}_i + \gamma_i \overline{C}_i \right\}, \quad (28)$$

### B. Solution:

The Bellman optimality equation for problem in (28) is given by:

$$V^*(\mathbf{Q}) = E \left[ \min_{\pi(\chi)} \left[ c(\pi, \chi) \tau + \mu \sum_{\mathbf{Q}'} pr(\mathbf{Q}' | \pi, \chi) V^*(\mathbf{Q}') \right] \middle| \mathbf{Q} \right], \quad (29)$$

Similar to previous section, we first solve the decomposed problem assuming zero-valued cross-link and then solve it up to its second (linear) Taylor term. The theorems providing the implicit equations for $J_i(Q_i)$ and their derivatives can be found in appendix D.

## VII. SIMULATIONS & RESULTS

### A. Jointly optimal allocations of fronthaul and power

Simulations are conducted for a cooperation cluster of size two. Large-scale fading is constant during each simulation run. The small-scale fading has a Rayleigh distribution with a mean equal to the large-scale fading, and is updated at each time slot. Several simulation runs were conducted with various large-scale fading matrices. The off-diagonal entries of these matrices were chosen one order of magnitude smaller than the diagonals. Fig. 2 compares the performance of the joint allocation and an extension of the fixed power method in [6] for their sum of average delays over various arrival rates of one of the users. In order to have a fair comparison, the average of users' powers over time in the proposed method was set equal to those in the fixed-power scheme. The wireless channel bandwidth is 2 MHz and the total fronthaul capacity is set to 10 bits/sec/Hz.

We see that adding dynamic *power control* to the decisions set has resulted in considerable reduction in the sum of delays. The point labeled *inf* is actually the instability point of the fixed-power method. Overall, the gain in delay reduction increases as the sum of users' data rates increases, and the system becomes more crowded.

Joint allocation has also extended the stability region of the arrival rates by limiting the average delay for cases in which the previous method went instable. Fluctuations in Fig.2 for the *constant-power* method, is due to *step-like* operation of the priority function $J(\mathbf{Q}; \delta)$.

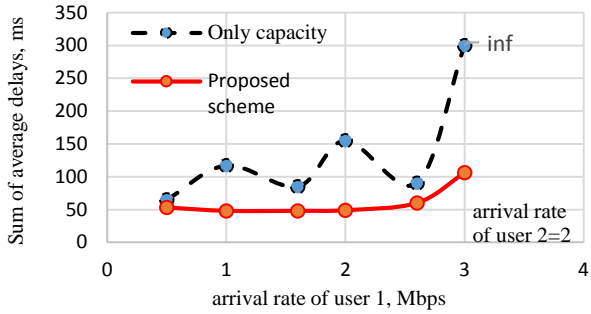

Fig. 2. Sum of the users' average delays for various arrival rates of user 1, for the proposed and earlier methods

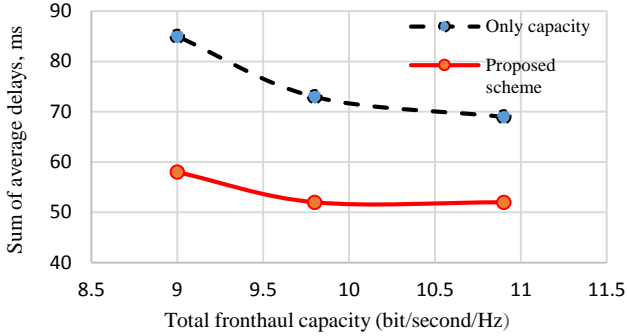

Fig. 3. Sum of the users' average delays as functions of total fronthaul capacity for the proposed and previous methods

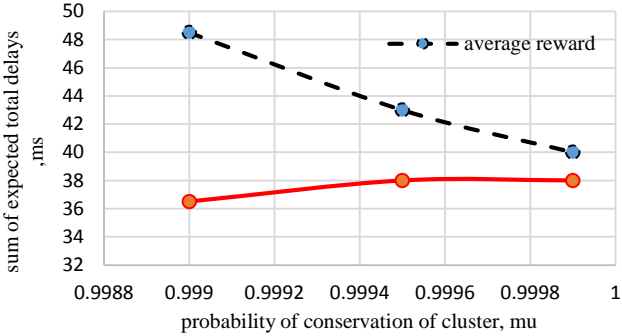

Fig. 4: Sum of the expected total delays of the users as functions of $\mu$, for the proposed and custom methods.

Therefore, the next advantage of the proposed scheme is its smoother behavior. Fig. 3 compares the sum of average delays for the proposed and fixed-power methods as functions of the total fronthaul capacity, considering a constant pair of arrival rates and considerable reduction of delay is obvious here as well. The gain in delay reduction increases as the total fronthaul capacity decreases or equivalently, the network becomes more crowded.

### B. Finite-duration services

In this section, we simulated a CRAN with parameters same as fixed-power model in previous section, but over a random simulation time (horizon) with geometrical distribution of parameter $\mu$, and examined the performance of the two algorithms; the proposed discounted-reward and the custom average-reward schemes. For each $\mu$, the simulations were repeated 50 times to account for realizations of various optimization lengths and averaged over them. Fig. 4 shows sum of the accumulated delays of users, averaged over 50 trials, for the two methods.

Fig. 4 shows that as $\mu$ decreases (or as departure of at least one of the users becomes more probable), the proposed scheme results in greater gain over the average-reward model and creates much less total delay for the users, on average. On the other hand, as $\mu$ increases, the users become more likely to ask for longer services and the result of the proposed solution approaches to the previous one as expected.

## VIII. CONCLUSIONS

We considered the problem of minimizing users' delays in uplink of a CRAN, which is a multi-objective optimization. Primarily, we chose the average delay criteria for each user, defined as an infinite-horizon average-reward objective function. Combining the MDP and Pareto optimization theories, we proved that every possible balance in optimizing a number of infinite-horizon average-reward objective functions is achievable by optimizing a properly weighted sum of those objective functions. Consequently, every balance point in minimizing users' average delays is also achievable by minimizing a properly weighted sum of average delays. Moreover, we analyzed the mechanism through which, *dynamic control* of transmit powers can improve the effectiveness of fronthaul allocation, in terms of reducing users' average delays. The proposed policy results in significantly lower weighted sum delay in comparison with the *fixed-power* scheme and also extends the stability region of the arrival rates. Finally, we analyzed the more realistic scenario of finite service durations for the users. In this scenario, the horizon becomes finite random value with a geometrical distribution. Consequently we were able to choose the more preferred, *total delay* criterion instead of the average-delay. The proposed scheme significantly outperforms the average-reward model in total delay reduction, especially when departure or entrance of a user is more probable at each time slot.

**Appendix A:** (proof of theorem 5)

Assuming $\delta = 0$, we have $R_i = R_i^0 = \dfrac{W}{2}\log_2\left(1 + \dfrac{p_i}{\|S_{ii}\|^2(N_i+\sigma^2)}\right)$ and $N_i = \dfrac{p_i\|H_{ii}\|^2+\sigma^2}{2^{C_i}-1}$.

By dismissing cross-links, data flows of different users become independent and hence we can write equation (22) for a network of only one data flow:

$$E\left[\min_{C_i,P_{i,d}}\left[\beta_i\dfrac{Q_i}{\lambda_i}+\gamma C_i+\mu_i P_{i,d}+\dfrac{\partial J_i(Q_i)}{\partial Q_i}\left(\lambda_i - R_i^0\left(H_{ii},C_i,P_{d,i}\right)\right)\right]\bigg| Q_i\right] - c_i^\infty = 0 \quad (30)$$

We denote the objective function of the optimization in (30), as $u$. If the optimum is an interior point, the gradient of $u$ must be zero.

Setting derivative of $u$ with respect to $C_i$ equal to zero, results in:

$$2^{C_i^*} = \dfrac{\left(p_0+p_{i,d}^*\right)\|H_{ii}\|^2}{\sigma^2}\left(\dfrac{WJ_i'(Q_i)}{2\gamma}-1\right)^+ \quad (31)$$

Defining some intermediate variables:

$$x_i \equiv \dfrac{P_{i,d}\|H_{ii}\|^2}{\sigma^2},\; x_{0i} \equiv \dfrac{P_{i,0}\|H_{ii}\|^2}{\sigma^2},\; y_i \equiv 2^{C_i},\; \alpha_i \equiv \dfrac{WJ_i'(Q_i)}{2\gamma} \quad (32)$$

we can write (32) in a simpler form:

$$y_i^* = (x_{0i}+x_i^*)(\alpha_i-1)^+ \quad (33)$$

Furthermore, the derivative of $u$ with respect to dynamic power must also be zero:

$$\dfrac{\partial u}{\partial x_i} = \mu_i\dfrac{\sigma^2}{\|H_{ii}\|^2} - \dfrac{WJ_i'}{2\ln 2}\times\dfrac{y_i-1}{(x_i+x_{0i}+1)(x_i+x_{0i}+y_i)} = 0 \quad (34)$$

By defining $\tilde{k}_i \equiv \dfrac{2\ln(2)\mu_i\sigma^2}{\|H_{ii}\|^2 WJ_i'}$, we can rewrite (34) as:

$$\dfrac{y_i^*-1}{(x_i^*+x_{0i}+1)(x_i^*+x_{0i}+y_i^*)} = \tilde{k}_i \quad (35)$$

Equations (34) and (36) constitute a system of two non-linear equations. Substituting (33) in (35):

$$\dfrac{(x_{0i}+x_i^*)(\alpha_i-1)-1}{(x_i^*+x_{0i}+1)(x_i^*+x_{0i})\alpha_i} = \tilde{k}_i \quad (36)$$

Defining $k_i^{-1} \equiv \alpha_i\tilde{k}_i = \dfrac{\mu_i\sigma^2\ln 2}{\gamma_i\|H_{ii}\|^2}$ and $x = x_{0i}+x_i^*$, we may rewrite (36) as:

$$x^2+(1-k_i(\alpha_i-1))x+k_i = 0 \rightarrow$$

$$x = \dfrac{1}{2}\left(k_i(\alpha_i-1)-1\pm\sqrt{k_i^2(\alpha_i-1)^2+1-2k_i(\alpha_i+1)}\right) \quad (37)$$

Therefore:

$$x_i^* = \dfrac{1}{2}\left(k_i(\alpha_i-1)-1\pm\sqrt{k_i^2(\alpha_i-1)^2+1-2k_i(\alpha_i+1)}\right)-x_{0i} \quad (38)$$

The expression under the square-root must be non-negative. It is also a function of $\|H_{ii}\|^2$ due to the term $k_i$. Therefore, in the channel situations where it becomes negative, $x_i^* = 0$. Its roots as function of $k_i$ are:

$$\frac{\alpha_i + 1 \pm \sqrt{(\alpha_i+1)^2 - (\alpha_i-1)^2}}{(\alpha_i-1)^2} = \frac{\alpha_i + 1 \pm 2\sqrt{\alpha_i}}{(\alpha_i-1)^2} = \frac{(\sqrt{\alpha_i} \pm 1)^2}{(\alpha_i-1)^2} = \begin{cases} \frac{1}{(\sqrt{\alpha_i}-1)^2} \\ \frac{1}{(\sqrt{\alpha_i}+1)^2} \end{cases} \quad (39)$$

Between these two roots, the term under the square-root in (38) is negative. Thus, we must have $k_i > \frac{1}{(\sqrt{\alpha_i}-1)^2}$, or $k_i < \frac{1}{(\sqrt{\alpha_i}+1)^2}$. But in the second case $x < 0$. Consequently, only the value of $k_i > \frac{1}{(\sqrt{\alpha_i}-1)^2}$ is acceptable. By such constraint on $k_i$, the smaller root in (38) will be negative and therefore the only answer of equation in (37) is:

$$x_i^* = \begin{cases} \frac{1}{2}\left(k_i(\alpha_i-1) - 1 + \sqrt{k_i^2(\alpha_i-1)^2 + 1 - 2k_i(\alpha_i+1)}\right) - x_{0i} & k_i \geq \frac{1}{(\sqrt{\alpha_i}-1)^2} \\ 0 & O.W \end{cases} \quad (40)$$

Furthermore, the second necessary condition for an interior optimum is that the "Hessian" matrix of 'u' must be non-negative at the optimum point;

$$|H_u| = \frac{WJ_i'}{2\ln 2} \times \frac{(y_i-1)(2x_i + 2x_{0i} + y_i + 1)}{(x_i + x_{0i} + 1)^2 (x_i + x_{0i} + y_i)^2} \propto \frac{(x_i + x_{0i} + y_i)}{(x_i + x_{0i} + 1)^2} \times \left((x_i + x_{0i})(y_i - 2) - 1\right) \quad (41)$$

Therefore, we can rewrite the second necessary condition as:

$$\left((x_i^* + x_{0i})(y_i^* - 2) - 1\right) \geq 0 \quad (42)$$

Substituting $y_i^*$ in terms of $x_i^*$, in (42):

$$(\alpha_i - 1)x^2 - 2x - 1 \geq 0 \rightarrow x = x_i^* + x_{0i} > \frac{1}{\sqrt{\alpha_i}-1} \quad (43)$$

Replacing $x_i^*$ from (40) in (42), we obtain the same condition of $k_i > \frac{1}{(\sqrt{\alpha_i}-1)^2}$.

Consequently, the optimal policy in case of no cross-links, is given by (40).

The condition $k_i > \frac{1}{(\sqrt{\alpha_i}-1)^2}$ is equivalent to the condition (44) on $\|\tilde{H}_{ii}\|^2$:

$$\frac{\gamma L_{ii} \|\tilde{H}_{ii}\|^2}{\mu_i \sigma^2 \ln 2} > \frac{1}{(\sqrt{\alpha_i}-1)^2} \rightarrow \|\tilde{H}_{ii}\|^2 > \frac{\mu_i \sigma^2 \ln 2}{\gamma L_{ii} (\sqrt{\alpha_i}-1)^2} \equiv h_{i0} \quad (44)$$

In order to calculate $E\left[\gamma C_i^*\right]$ and $E\left[\mu_i P_{i,d}^*\right]$, we define $z \equiv \|\tilde{H}_{ii}\|^2$ and obtain:

$$E\left[\mu_i P_{i,d}^*\right] = E\left[\mu_i x_i^* \frac{\sigma^2}{\|H_{ii}\|^2}\right] = \int_{h_{i0}}^{\infty} \frac{\mu_i \sigma^2}{L_{ii}} \frac{x_i^*(z)}{z} e^{-z} dz \quad (45)$$

We also define $b_i \equiv \frac{\gamma L_{ii}}{\mu_i \sigma^2 \ln 2}$. Therefore, $k_i = b_i z$ and:

$$x_i^*(z) = \frac{1}{2}\left(b_i z(\alpha_i - 1) - 1 + \sqrt{b_i^2 z^2 (\alpha_i - 1)^2 + 1 - 2b_i z(\alpha_i + 1)}\right) - x_{0i} \quad (45)$$

Substituting (46) in (45), the integral would not have a closed-form solution. However, for sufficiently large $\alpha_i$ values, the square-root term can be approximated by $b_i z(\alpha_i - 1) - 1$ and hence:

$$x_i^*(z) \approx b_i z(\alpha_i - 1) - 1 - \frac{P_{0i} L_{ii}}{\sigma^2} z \quad (47)$$

$$E\left[\mu_i P_{i,d}^*\right] \approx \int_{h_{i0}}^{\infty} \frac{\mu_i \sigma^2}{L_{ii}} \frac{\left(b_i z(\alpha_i - 1) - 1 - \frac{P_{0i} L_{ii}}{\sigma^2} z\right)}{z} e^{-z} dz = \left(\frac{\gamma}{\ln 2}(\alpha_i - 1) - \mu_i P_{0i}\right) e^{-h_{i0}} - \frac{\mu_i \sigma^2}{L_{ii}} E_1(h_{i0}) \quad (48)$$

We will also use the approximation (47) for calculating $E\left[C_i^*\right]$. On the other hand, we must remember that for $z < h_{i0}$, the transmit power is $P_{0i}$ and as long as $x_{0i}(\alpha_i - 1)^+ > 1$, the capacity allocated to user 'i' and its data rate are not zero. This constraint is equivalent to constraint (49) on $\|\tilde{H}_{ii}\|^2$:

$$\frac{P_{0i} L_{ii} \|\tilde{H}_{ii}\|^2}{\sigma^2}(\alpha_i - 1)^+ > 1 \rightarrow \|\tilde{H}_{ii}\|^2 > \frac{\sigma^2}{P_{0i} L_{ii}(\alpha_i - 1)^+} \equiv h_{i4}, \quad h_{i3} \equiv \min\{h_{i0}, h_{i4}\} \quad (49)$$

$$E\left[C_i^*\right] = E\left[\log_2\left((x_{0i} + x_i^*)(\alpha_i - 1)\right)\right] \approx \int_{h_{i3}}^{h_{i0}} \log_2\left((\alpha_i - 1)\frac{P_{0i} L_{ii} z}{\sigma^2}\right) e^{-z} dz +$$

$$\int_{h_{i0}}^{\infty} \log_2\left((\alpha_i - 1)(b_i z(\alpha_i - 1) - 1)\right) e^{-z} dz \quad (50)$$

$$E\left[C_i^*\right] \approx -e^{-h_{i0}} \log_2\left(\frac{h_{i0}}{h_{i3}}\right) + \frac{1}{\ln 2}\left(E_1(h_{i3}) - E_1(h_{i0})\right) +$$

$$e^{-h_{i0}} \log_2\left((\alpha_i - 1)(b_i h_{i0}(\alpha_i - 1) - 1)\right) + \frac{1}{\ln 2} e^{-\frac{1}{b_i(\alpha_i - 1)}} E_1\left(h_{i0} - \frac{1}{b_i(\alpha_i - 1)}\right) \quad (51)$$

Next, we should calculate $R_{i0}$.

$$N_i^* + \sigma^2 = \frac{p_i^* \|H_{ii}\|^2 + \sigma^2 2^{C_i^*}}{2^{C_i^*} - 1} = \frac{\sigma^2\left(y_i^* + x_{0i} + x_i^*\right)}{y_i^* - 1} = \frac{\sigma^2\left(x_{0i} + x_i^*\right)\alpha_i}{y_i^* - 1} \quad (52)$$

$$R_{i0}^* = \frac{W}{2} \log_2\left(1 + \frac{p_i^* \|H_{ii}\|^2}{(N_i^* + \sigma^2)}\right) = \frac{W}{2} \log_2\left(1 + \frac{p_i^* \|H_{ii}\|^2}{\frac{\sigma^2(x_{0i} + x_i^*)\alpha_i}{y_i^* - 1}}\right) = \frac{W}{2} \log_2\left(1 + \frac{y_i^* - 1}{\alpha_i}\right) \quad (53)$$

$$\frac{2}{W} E\left[R_{i0}^*\right] = E\left[\log_2\left(1 + \frac{y_i^* - 1}{\alpha_i}\right)\right] = E\left[\log_2\left(1 + \frac{(x_{0i} + x_i^*)(\alpha_i - 1) - 1}{\alpha_i}\right)\right] \approx$$

$$\int_{h_{i3}}^{h_{i0}} \log_2\left(1 + \frac{\frac{z}{h_{i3}} - 1}{\alpha_i}\right) e^{-z} dz + \int_{h_{i0}}^{\infty} \log_2\left(1 + \frac{(b_i z(\alpha_i - 1) - 1)(\alpha_i - 1) - 1}{\alpha_i}\right) e^{-z} dz = \quad (54)$$

$$\int_{h_{i3}}^{h_{i0}} \log_2\left(\frac{z}{\alpha_i h_{i3}} + \frac{\alpha_i - 1}{\alpha_i}\right) e^{-z} dz + \int_{h_{i0}}^{\infty} \log_2\left(\frac{(\alpha_i - 1)^2 b_i z}{\alpha_i}\right) e^{-z} dz$$

$$E\left[R_{i0}^{*}\right] = \frac{W}{2}(-e^{-h_{i0}}\log_2\left(\frac{h_{i0}}{\alpha_i h_{i3}}+\frac{\alpha_i-1}{\alpha_i}\right)+\frac{e^{h_{i3}(\alpha_i-1)}}{\ln 2}\left[E_1(h_{i3}\alpha_i)-E_1(h_{i0}+h_{i3}(\alpha_i-1))\right]+$$
$$e^{-h_{i0}}\log_2\left(\frac{(\alpha_i-1)^2 b_i h_{i0}}{\alpha_i}\right)+\frac{1}{\ln 2}E_1(h_{i0}))$$
(55)

The proof of theorem 4 is completed here.

## Appendix B: (proof of theorem 6)

Proof:

Substituting (22) in (21) and computing its derivative with respect to $L_{ij}$:

$$\tilde{\theta}_{ij}\equiv\frac{\partial\theta}{\partial L_{ij}}=\sum_{k=1}^{m}\frac{\partial\tilde{J}_{ij}(\mathbf{Q})}{\partial Q_k}\left(\lambda_k-E\left[R_k^0\right]\right)-\sum_{k=1}^{m}J_k'(Q_k)E\left[\frac{\partial R_k}{\partial L_{ij}}\right] \quad (56)$$

$$\frac{\partial R_k}{\partial L_{ij}}=\frac{-W}{2}\frac{p_i}{\left(p_i+\sum_{r=1}^{L}\|S_{kr}\|^2(N_r^*+\sigma^2)\right)}\frac{1}{\sum_{r=1}^{L}\|S_{kr}\|^2(N_r^*+\sigma^2)}\frac{\partial\sum_{r=1}^{m}\|S_{kr}\|^2(N_r^*+\sigma^2)}{\partial L_{ij}} \quad (57)$$

$$\frac{\partial N_r^*}{\partial L_{ij}}=\begin{cases}\frac{p_j\|\tilde{H}_{ij}\|^2}{2^{C_i}-1} & r=i \\ 0 & r\neq i\end{cases}, \quad \frac{\partial\|S_{kr}\|^2}{\partial L_{ij}}=\begin{cases}\frac{\|\tilde{H}_{ij}\|^2}{\|H_{ii}\|^2\|H_{jj}\|^2} & k=i, r=j \\ 0 & O.W\end{cases} \quad (58)$$

Replacing (58) in (57), we have:

$$\frac{\partial R_k}{\partial L_{ij}}=\frac{-W}{2}\frac{1}{\left(1+\frac{(N_i^*+\sigma^2)}{p_i^*\|H_{ii}\|^2}\right)}\frac{\|\tilde{H}_{ij}\|^2}{(N_i^*+\sigma^2)}\left\{\frac{p_j^*}{2^{C_i}-1}+\frac{(N_j^*+\sigma^2)}{\|H_{jj}\|^2}\right\}\delta(k,i) \quad (59)$$

Using (33), (40), (47) and (52), we can rewrite (59) in a simpler form:

$$\frac{\partial R_k}{\partial L_{ij}}=\frac{-W}{2}\frac{1}{\left(1+\frac{\alpha_i}{(y_i^*-1)}\right)}\frac{\|\tilde{H}_{ij}\|^2(y_i^*-1)}{\sigma^2(x_{0i}+x_i^*)\alpha_i}\left\{\frac{\sigma^2}{\|H_{jj}\|^2}\frac{(x_{0j}+x_j^*)}{y_i^*-1}+\frac{\frac{\sigma^2(x_{0j}+x_j^*)\alpha_j}{y_j^*-1}}{\|H_{jj}\|^2}\right\}\delta(k,i)=$$
$$\frac{-W}{2}\frac{1}{\left(1+\frac{\alpha_i}{(y_i^*-1)}\right)}\frac{\|\tilde{H}_{ij}\|^2}{\|H_{jj}\|^2\alpha_i}\frac{(x_{0j}+x_j^*)}{(x_{0i}+x_i^*)}\left\{1+\frac{\alpha_j(y_i^*-1)}{y_j^*-1}\right\}\delta(k,i)$$
(60)

Considering independency of elements of **H**, we can write:

$$E\left[\frac{\partial R_i}{\partial L_{ij}}\right]=\frac{-W}{2}\frac{E\left[\|\tilde{H}_{ij}\|^2\right]}{L_{ij}\alpha_i}\left\{E\left[\frac{(y_i^*-1)}{(x_{0i}+x_i^*)(y_i^*-1+\alpha_i)}\right]E\left[\frac{(x_{0j}+x_j^*)}{\|\tilde{H}_{jj}\|^2}\right]+\alpha_j E\left[\frac{(y_i^*-1)^2}{(x_{0i}+x_i^*)(y_i^*-1+\alpha_i)}\right]E\left[\frac{1}{\|\tilde{H}_{jj}\|^2}\frac{(x_{0j}+x_j^*)}{y_j^*-1}\right]\right\} \quad (61)$$

If we have enough fronthaul capacity, $y_j^*-1\approx y_j^*$. Combining this with (39), we can simplify (61) further to:

$$E\left[\frac{(y_i^*-1)}{(x_{0i}+x_i^*)(y_i^*-1+\alpha_i)}\right]\approx E\left[\frac{(x_{0i}+x_i^*)(\alpha_i-1)}{(x_{0i}+x_i^*)(\alpha_i-1)(x_{0i}+x_i^*+1)}\right]=E\left[\frac{1}{(x_{0i}+x_i^*+1)}\right]\approx\xrightarrow{a_i=\frac{\sigma^2}{P_{0i}L_{ii}}}$$
$$\int_{h_{i3}}^{h_{i0}}\frac{e^{-z}}{a_i^{-1}z+1}dz+\int_{h_{i0}}^{\infty}\frac{e^{-z}}{b_i z(\alpha_i-1)}dz=a_i e^{a_i}\left(E_1(h_{i3}+a_i)-E_1(h_{i0}+a_i)\right)+\frac{E_1(h_{i0})}{b_i(\alpha_i-1)}$$
(62)

$$E\left[\frac{(y_i^*-1)^2}{(x_{0i}+x_i^*)(y_i^*-1+\alpha_i)}\right] \approx E\left[\frac{(x_{0i}+x_i^*)^2(\alpha_i-1)^2}{(x_{0i}+x_i^*)(\alpha_i-1)(x_{0i}+x_i^*+1)}\right] = (\alpha_i-1)\left\{1-E\left[\frac{1}{x_{0i}+x_i^*+1}\right]\right\} \quad (63)$$

$$E\left[\frac{(x_{0j}+x_j^*)}{\|\tilde{H}_{jj}\|^2}\right] = \int_{h_{j3}}^{h_{j0}} \frac{P_{0j}L_{jj}}{\sigma^2}e^{-z}dz + \int_{h_{j0}}^{\infty}\left(b_j(\alpha_j-1)-\frac{1}{z}\right)e^{-z}dz = \qquad (64)$$

$$\frac{e^{-h_{j3}}-e^{-h_{j0}}}{a_j} + b_j(\alpha_j-1)e^{-h_{j0}} - E_1(h_{j0})$$

$$E\left[\frac{1}{\|\tilde{H}_{jj}\|^2}\frac{(x_{0j}+x_j^*)}{y_j^*-1}\right] \approx E\left[\frac{1}{\|\tilde{H}_{jj}\|^2}\frac{(x_{0j}+x_j^*)}{(x_{0j}+x_j^*)(\alpha_j-1)}\right] = \frac{E_1[h_{j3}]}{(\alpha_j-1)} \quad (65)$$

Substitution of these terms in (56) results in a complicated partial differential equation for which there is no closed-form solution. Assuming big-enough $\alpha_i$ s:

$$E\left[\frac{(y_i^*-1)}{(x_{0i}+x_i^*)(y_i^*-1+\alpha_i)}\right] \approx v_1 + \frac{v_2}{(\alpha_i-1)}, \quad E\left[\frac{(y_i^*-1)^2}{(x_{0i}+x_i^*)(y_i^*-1+\alpha_i)}\right] \approx (\alpha_i-1)(1-v_1)-v_2$$

$$E\left[\frac{(x_{0j}+x_j^*)}{\|\tilde{H}_{jj}\|^2}\right] = v_3 + v_4(\alpha_j-1), \quad E\left[\frac{1}{\|\tilde{H}_{jj}\|^2}\frac{(x_{0j}+x_j^*)}{y_j^*-1}\right] \approx \frac{v_5}{(\alpha_j-1)} \qquad (66)$$

where, the coefficients $v_i$ can be derived from equations (62) to (65).
Substituting these in (56):

$$\frac{\partial \tilde{J}_{ij}}{\partial Q_i}\left(E[R_{i0}^*]-\lambda_i\right) + \frac{\partial \tilde{J}_{ij}}{\partial Q_j}\left(E[R_{j0}^*]-\lambda_j\right) = \frac{\gamma_i}{L_{jj}}\left\{\left(v_1+\frac{v_2}{\alpha_i}\right)(v_3+v_4\alpha_j)+v_5\left((1-v_1)\alpha_i-v_2\right)\right\} \quad (67)$$

With approximation of large values of $\alpha_i$, we can neglect constant terms and terms with $\alpha_k$ in their denominator and achieve:

$$\frac{\partial u_1}{\partial Q_i}\left(E[R_{i0}^*]-\lambda_i\right) + \frac{\partial u_1}{\partial Q_j}\left(E[R_{j0}^*]-\lambda_j\right) = \frac{\gamma_i}{L_{jj}}\left(v_5(1-v_1)\alpha_i + v_1v_4\alpha_j\right) \quad (68)$$

The answer to this PDE is:

$$\frac{\partial u_1}{\partial Q_i} = \frac{\gamma_i}{L_{jj}}\frac{v_5(1-v_1)\alpha_i}{\left(E[R_{i0}^*]-\lambda_i\right)}, \quad \frac{\partial u_1}{\partial Q_j} = \frac{\gamma_i}{L_{jj}}\frac{v_1v_4\alpha_j}{\left(E[R_{j0}^*]-\lambda_j\right)} \quad (69)$$

## Appendix C: (proof of lemma 2)

let $r(t)$ be the instantaneous reward at time slot $t$ upon being in state $\chi(t)$ and taking the action y(t), and $R_T$ be the total reward over a duration of T. Then, the expected total reward is equal with:

$$E[R] = E^\pi[E_T[R|T]] = E^\pi\left[\sum_{T=1}^{\infty}\left(\sum_{t=1}^{T}r(\chi(t),y(t))\right)(1-\mu)\mu^{T-1}\right] = E^\pi\left[\sum_{t=1}^{\infty}\sum_{T=1}^{\infty}r(\chi(t),y(t))(1-\mu)\mu^{T-1}\right] =$$

$$E^\pi\left[\sum_{t=1}^{\infty}r(\chi(t),y(t))\mu^{t-1}\right], \quad (70)$$

## Appendix D: (theorems with proofs for finite service time scenario)

**Theorem 8**:
*Assume there exists J (Q; δ) of class $C^2(\mathbb{R}_+^n)$ which satisfies:*

1) *The following partial differential equation:*

$$0 = \left[ \min_{\pi(\chi)} \left[ c(\pi,\chi) + \mu \sum_{i=1}^{n} \left( \frac{\partial J^*(\mathbf{Q})}{\partial Q_i} (\lambda_i - E[R_i(\mathbf{H},C)]) \right) \right] \middle| \mathbf{Q} \right], \qquad (71)$$

with the boundary condition: $J(0;\delta) = 0$

2) For all $k$, $\frac{\partial J(\mathbf{Q};\delta)}{\partial Q_k}$ is an increasing function of all $Q_k$.

3) $J(\mathbf{Q};\delta) = O(\|\mathbf{Q}\|^2)$

Then, $V^*(\mathbf{Q}) = J(\mathbf{Q};\delta) + o(1)$ and the error term $o(1)$ tends asymptotically to zero for sufficiently small $\tau$.

*Proof:*

In (40), $\sum_{\mathbf{Q}'} pr(\mathbf{Q}'|\chi,\Omega(\chi)) V^*(\mathbf{Q}')$ is equivalent with $E[V^*(\mathbf{Q}')|\mathbf{Q}]$. Moreover we know that

$Q_i' = Q_i - R_i(\mathbf{H},C)\tau + A_i\tau$. Therefore, we can write the Taylor expansion of $V(\mathbf{Q}')$ as:

$$E[V(\mathbf{Q}')|\mathbf{Q}] = V(\mathbf{Q}) + \sum_{i=1}^{L} \left( \frac{\partial V(\mathbf{Q})}{\partial Q_i} (\lambda_i - E[R_i(\mathbf{H},C)]) \right) \tau + o(\tau) \qquad (72)$$

Substituting (72) in (29), we will have:

$$V^*(\mathbf{Q}) = E\left[ \min_{\Omega(\chi)} \left[ c(\mathbf{Q},\Omega(\chi))\tau + V^*(\mathbf{Q}) + \mu \sum_{i=1}^{L} \left( \frac{\partial V^*(\mathbf{Q})}{\partial Q_i} (\lambda_i - E[R_i(\mathbf{H},C)]) \right) \tau + o(\tau) \right] \middle| \mathbf{Q} \right] \qquad (73)$$

By omitting $V^*(\mathbf{Q})$ from the two sides of (73) and dividing both by $\tau$ we achieve:

$$0 = E\left[ \min_{\Omega(\chi)} \left[ c(\mathbf{Q},\Omega(\chi)) + \mu \sum_{i=1}^{L} \left( \frac{\partial V^*(\mathbf{Q})}{\partial Q_i} (\lambda_i - E[R_i(\mathbf{H},C)]) \right) + \frac{o(\tau)}{\tau} \right] \middle| \mathbf{Q} \right] \qquad (74)$$

Comparing (74) with (71), the assertion of theorem 8 becomes obvious. The next two conditions are to ensure the admissibility of the policy and are proved in the same way as in [8].

**Theorem 9:**

*The approximate priority function is decomposable into sub-priority functions $J_i(Q_i)$:*

$$J(\mathbf{Q};0) = \sum_{i=1}^{n} J_i(Q_i) , \qquad (75)$$

and the $J_i(Q_i)$ for every $i$, approves the following implicit equation:

$$\beta_i \frac{Q_i}{\lambda_i} + \frac{\gamma_i}{\ln 2} E_1 \left( \frac{a_i \gamma_i}{\left( \mu \frac{W}{2} J_i'(Q_i) - \gamma_i \right)} \right) + \mu J_i'(Q_i) \lambda_i - \mu \frac{W}{2} J_i'(Q_i) \frac{e^{a_i}}{\ln 2} E_1 \left( \frac{a_i \mu \frac{W}{2} J_i'(Q_i)}{\left( \mu \frac{W}{2} J_i'(Q_i) - \gamma_i \right)} \right) = 0 . \qquad (76)$$

In which, $a_i = \frac{\sigma^2}{pL_{ii}}$ and $E_1(z) \equiv \int_1^\infty \frac{e^{-tz}}{t} dt = \int_z^\infty \frac{e^{-t}}{t} dt$.

*Proof:* first we prove the decomposability:

Assuming $\delta = 0$, $R_i = R_i^0 = \frac{W}{2} \log_2 \left( 1 + \frac{p}{\|S_{ii}\|^2 (N_i + \sigma^2)} \right)$ and $N_i = \frac{p\|H_{ii}\|^2 + \sigma^2}{2^{C_i} - 1}$, the queues will evolve independently and the equation (71) for a single user is given by:

$$E\left[\min_{C_i}\left[\beta_i\frac{Q_i}{\lambda_i}+\gamma_i C_i+\mu\frac{\partial J_i(Q_i)}{\partial Q_i}\left(\lambda_i-R_i^0(H_{ii},C_i)\right)\right]\Big|Q_i\right]=0 \quad . \tag{77}$$

The minimization in (77) can be solved easily and by setting its derivative equal with zero. The results are:

$$C_i^* = \left(\log_2\left(\frac{p\|H_{ii}\|^2}{\sigma^2}\left(\frac{\mu\frac{W}{2}J_i'(Q_i)}{\gamma_i}-1\right)^+\right)\right)^+ \tag{78}$$

$$E\left[\gamma_i C_i^*|Q_i\right] = \begin{cases} \dfrac{\gamma_i}{\ln 2}E_1\left(\dfrac{\sigma^2\gamma_i}{pL_{ii}\left(\mu\frac{W}{2}J_i'(Q_i)-\gamma_i\right)}\right) & \mu\frac{W}{2}J_i'(Q_i)>\gamma_i \\ 0 & \mu\frac{W}{2}J_i'(Q_i)\le\gamma_i \end{cases} \tag{79}$$

$$E\left[R_i^0(H_{ii},C_i)|Q_i\right] = \begin{cases} \dfrac{W}{2}\dfrac{e^{\frac{\sigma^2}{pL_{ii}}}}{\ln 2}E_1\left(\dfrac{\sigma^2\mu\frac{W}{2}J_i'(Q_i)}{pL_{ii}\left(\mu\frac{W}{2}J_i'(Q_i)-\gamma_i\right)}\right) & \mu\frac{W}{2}J_i'(Q_i)>\gamma_i \\ 0 & \mu\frac{W}{2}J_i'(Q_i)\le\gamma_i \end{cases} \tag{80}$$

Substituting (79) and (80) in (71), we achieve (75).
**Theorem 10:**
*The second Taylor term of the priority function* $J(\mathbf{Q};\delta)$ *in terms of* $L_{ij}$ *s is given by:*

$$\frac{\partial J(\mathbf{Q};\delta)}{\partial Q_i} = J_i'(Q_i)\left(1+\frac{2}{L_{jj}}\frac{\left(1-a_i e^{a_i}E_1(a_i)\right)}{\dfrac{e^{a_i}E_1(a_i)}{\ln 2}-\dfrac{2\lambda_i}{W}}\right) \quad . \tag{81}$$

Proof:
The second-order Taylor expansion of $J(\mathbf{Q};\delta)$ with respect to $L_{ij}$ s is given by:

$$J(\mathbf{Q};\delta) = J(\mathbf{Q};0)+\sum_{i\ne j}\sum_{j=1}^{L}L_{ij}\tilde{J}_{ij}(\mathbf{Q})+O(\delta^2) \quad . \tag{82}$$

And we want to find $\tilde{J}_{ij}$ s. Substituting (82) in (29) and taking derivatives with respect to $L_{ij}$ s:

$$\frac{\partial\theta}{\partial L_{ij}} = \sum_{k=1}^{L}\frac{\partial\tilde{J}_{ij}(\mathbf{Q})}{\partial Q_k}\left(\lambda_k-E\left[R_k^0\right]\right)-\sum_{k=1}^{L}J_k'(Q_k)E\left[\frac{\partial R_k}{\partial L_{ij}}\right]=0 \quad . \tag{83}$$

$$\frac{\partial R_k}{\partial L_{ij}} = \frac{-W}{2}\frac{p}{\left(p+\sum_{m=1}^{L}\|S_{km}\|^2(N_m^*+\sigma^2)\right)}\frac{1}{\sum_{m=1}^{L}\|S_{km}\|^2(N_m^*+\sigma^2)}\frac{\partial\sum_{m=1}^{L}\|S_{km}\|^2(N_m^*+\sigma^2)}{\partial L_{ij}} \quad . \tag{84}$$

$$\frac{\partial N_m^*}{\partial L_{ij}} = \begin{cases} \dfrac{p\|\tilde{H}_{ij}\|^2}{2^{C_i}-1} & m=i \\ 0 & m \neq i \end{cases} \qquad (85)$$

Based on the first-order Taylor expansion with respect to $L_{ij}$ s:

$$\frac{\partial \|S_{km}\|^2}{\partial L_{ij}} = \begin{cases} \dfrac{\|\tilde{H}_{ij}\|^2}{\|H_{ii}\|^2 \|H_{jj}\|^2} & k=i, m=j \\ 0 & O.W \end{cases} \qquad (86)$$

Substituting (85) and (86) in (84), we have:

$$\frac{\partial R_k}{\partial L_{ij}} = \frac{-W}{2} \frac{1}{\left(1+\dfrac{(N_i^*+\sigma^2)}{p\|H_{ii}\|^2}\right)} \frac{\|\tilde{H}_{ij}\|^2}{(N_i^*+\sigma^2)} \left\{\frac{p}{2^{C_i}-1} + \frac{(N_j^*+\sigma^2)}{\|H_{jj}\|^2}\right\} \delta(k,i) \qquad (87)$$

We will also utilize this relation in the proceeding in order to make equations simpler:

$$\frac{1}{\left(1+\dfrac{(N_i^*+\sigma^2)}{p\|H_{ii}\|^2}\right)} = \frac{p\|H_{ii}\|^2}{2^{C_i} N_i^*} = \frac{\sigma^2}{\left(\dfrac{\mu W J_i'}{2\gamma_i}-1\right)} \frac{1}{N_i^*} \qquad (88)$$

Now, we must calculate $E\left[\dfrac{\partial R_k}{\partial L_{ij}}\right]$. Since the elements of **H** are independent:

$$E\left[\frac{\partial R_i}{\partial L_{ij}}\right] = \frac{-W \cdot E\left[\|\tilde{H}_{ij}\|^2\right]}{2} \left\{\frac{p\sigma^2}{\left(\dfrac{\mu W J_i'}{2\gamma_i}-1\right)} E\left[\frac{1}{(N_i^*+\sigma^2)(pL_{ii}\|\tilde{H}_{ii}\|^2+\sigma^2)}\right] + E\left[\frac{\dfrac{1}{(N_i^*+\sigma^2)}}{\left(1+\dfrac{(N_i^*+\sigma^2)}{pL_{ii}\|\tilde{H}_{ii}\|^2}\right)}\right] \cdot E\left[\frac{(N_j^*+\sigma^2)}{L_{jj}\|\tilde{H}_{jj}\|^2}\right]\right\} \qquad (89)$$

By defining $\alpha_i \equiv \dfrac{\mu W J_i'}{2\gamma_i}$ and for large $\alpha_i$ s:

$$E\left[\frac{\dfrac{1}{(N_i^*+\sigma^2)}}{\left(1+\dfrac{(N_i^*+\sigma^2)}{pL_{ii}\|\tilde{H}_{ii}\|^2}\right)}\right] = \frac{1-a_i e^{a_i} E_1(a_i)}{\sigma^2} \quad, \quad E\left[\frac{(N_j^*+\sigma^2)}{L_{jj}\|\tilde{H}_{jj}\|^2}\right] = \frac{2\sigma^2}{L_{jj}} K \quad, \quad E\left[\|\tilde{H}_{ij}\|^2\right] = 1$$

$$E\left[\frac{1}{(N_i^*+\sigma^2)(pL_{ii}\|\tilde{H}_{ii}\|^2+\sigma^2)}\right] = E\left[\frac{(\alpha_j-1)}{\alpha_j \sigma^2 pL_{ii}(\|\tilde{H}_{ii}\|^2+a_i)}\right] = \frac{e^{a_i} E_1(a_i)}{\sigma^2 pL_{ii}} \qquad (90)$$

Substituting these in (89):

$$E\left[\frac{\partial R_i}{\partial L_{ij}}\right] = \frac{-W}{2} \left\{\frac{e^{a_i} E_1(a_i)}{\alpha_i L_{ii}} + \frac{2K}{L_{jj}}\left(1-a_i e^{a_i} E_1(a_i)\right)\right\} \qquad (91)$$

Substituting (91) in (83):

$$\sum_{k=1}^{L} \frac{\partial \tilde{J}_{ij}(\mathbf{Q})}{\partial Q_k}\left(\lambda_k - \frac{We^{a_k} E_1(a_k)}{2\ln 2}\right) + \left\{\frac{\gamma_i e^{a_i} E_1(a_i)}{2L_{ii}} + \frac{WJ'_i}{2}\frac{2K}{L_{jj}}\left(1 - a_i e^{a_i} E_1(a_i)\right)\right\} = 0 \tag{92}$$

The first expression in the braces is a constant and in large $Q_i$ s, they can be ignored in comparison with $J'_i(Q_i) = O(Q_i)$
Therefore:

$$\frac{\partial \tilde{J}_{ij}(\mathbf{Q})}{\partial Q_i} \approx \frac{2J'_i(Q_i)}{L_{jj}} \frac{\left(1 - a_i e^{a_i} E_1(a_i)\right)}{\dfrac{e^{a_i} E_1(a_i)}{\ln 2} - \dfrac{2\lambda_i}{W}} \propto \frac{2Q_i}{L_{jj}} \frac{\left(1 - a_i e^{a_i} E_1(a_i)\right)}{\dfrac{e^{a_i} E_1(a_i)}{\ln 2} - \dfrac{2\lambda_i}{W}} \tag{93}$$